\documentclass[a4paper]{elsarticle}

\usepackage{amssymb}
\usepackage{hyperref}
\usepackage{lineno}

\usepackage{listings}
\usepackage{xcolor}

\definecolor{codegreen}{rgb}{0,0.6,0}
\definecolor{codegray}{rgb}{0.5,0.5,0.5}
\definecolor{codepurple}{rgb}{0.58,0,0.82}
\definecolor{backcolour}{rgb}{0.95,0.95,0.92}

\lstdefinestyle{mystyle}{
    backgroundcolor=\color{backcolour},   
    commentstyle=\color{codegreen},
    keywordstyle=\color{black},
    numberstyle=\tiny\color{codegray},
    stringstyle=\color{codepurple},
    basicstyle=\ttfamily\footnotesize,
    breakatwhitespace=false,         
    breaklines=true,                 
    captionpos=b,                    
    keepspaces=true,                 
    numbers=none,                    
    numbersep=5pt,                  
    showspaces=false,                
    showstringspaces=false,
    showtabs=false,                  
    tabsize=2
}

\journal{SoftwareX}

\begin{document}
\renewcommand{\labelenumii}{\arabic{enumi}.\arabic{enumii}}

\begin{frontmatter}

%https://www.overleaf.com/learn/latex/Code_listing

%%%%% TITLE %%%%%%%%%%%%%%%%%%%%%%%%%%%%%%%%%%%%%%%%%%%%%%
\title{\lstinline[basicstyle=\ttfamily\Large] |QMol-grid|: A MATLAB package for quantum-mechanical simulations in atomic and molecular systems}

\author[lsu]{Fran\c{c}ois Mauger\corref{corref}}
\ead{fmauger@lsu.edu}

\author[cnrs]{Cristel Chandre}
\ead{cristel.chandre@cnrs.fr}

\address[lsu]{Department of Physics and Astronomy, Louisiana State University, Baton Rouge, Louisiana 70803, USA}

\address[cnrs]{CNRS, Aix Marseille Univ, I2M, 13009 Marseille, France}

\cortext[corref]{Corresponding author}

%%%%% ABSTRACT %%%%%%%%%%%%%%%%%%%%%%%%%%%%%%%%%%%%%%%%%%%%%%
\begin{abstract}
    The \lstinline[basicstyle=\ttfamily\normalsize] |QMol-grid| package provides a suite of routines for performing quantum-mechanical simulations in atomic and molecular systems, currently implemented in one spatial dimension. It supports ground- and excited-state calculations for the Schr\"{o}dinger equation, density-functional theory, and Hartree-Fock levels of theory as well as propagators for field-free and field-driven time-dependent Schr\"{o}dinger equation (TDSE) and real-time time-dependent density-functional theory (TDDFT), using symplectic-split schemes.
    The package is written using MATLAB's object-oriented features and handle classes. It is designed to facilitate access to the wave function(s) (TDSE) and the Kohn-Sham orbitals (TDDFT) within MATLAB's environment.
\end{abstract}

%%%%% KEYWORDS %%%%%%%%%%%%%%%%%%%%%%%%%%%%%%%%%%%%%%%%%%%%%%

\end{frontmatter}

%\linenumbers

%%%%% METADATA %%%%%%%%%%%%%%%%%%%%%%%%%%%%%%%%%%%%%%%%%%%%%%
\section*{Metadata}

\begin{table}
\begin{tabular}{|l|p{6.5cm}|p{6.5cm}|}
\hline
{Nr.} & {Code metadata description} & \\ %{Please fill in this column} \\

\hline
C1 & Current code version & 1.21 \\
\hline
C2 & Permanent link to code/repository used for this code version & \url{https://github.com/fmauger1/QMol-grid.git} \\
\hline
C3 & Permanent link to Reproducible Capsule &  N/A \\
\hline
C4 & Legal Code License   & BSD-2-Clause \\
\hline
C5 & Code versioning system used &  git \\
\hline
C6 & Software code languages, tools, and services used & MATLAB (R2022a or later)~\cite{MATLAB} \\
\hline
C7 & Compilation requirements, operating environments \& dependencies & none \\
\hline
C8 & If available Link to developer documentation/manual & \url{https://github.com/fmauger1/QMol-grid/wiki} \\
\hline
C9 & Support email for questions & \url{fmauger@lsu.edu} \\
\hline
\end{tabular}
\caption{\lstinline[basicstyle=\ttfamily\normalsize] |QMol-grid| metadata}
\label{codeMetadata} 
\end{table}

%%%%% MANUSCRIPT %%%%%%%%%%%%%%%%%%%%%%%%%%%%%%%%%%%%%%%%%%%%%%

%% Motivation and significance
\section{Motivation and significance}

\textit{Ab initio} quantum simulations of the electronic structure and dynamics in atoms and molecules play an important role in many fields of physics and chemistry. They have lead to the development of many computational packages. For instance, optimized packages like~\cite{GAMESS,Gaussian,NWChem,Octopus,quantum_espresso} allow for routine quantum calculations in a range of atomic, molecular, and solid-state systems, typically running on high-performance computer (HPC) systems.
Alternatively, the \lstinline[basicstyle=\ttfamily\normalsize] |QMol-grid| package has been developed in the context of ultrafast atomic, molecular, and optical (AMO) research~\cite{Mauger_2022,Mauger_2024}, with a focus on low-dimension atomic and molecular models, (i) to provide a test bed for quantum-mechanical simulations that can easily run on personal computers, including when considering molecular systems with multiple interacting electrons and (ii) to facilitate access to the wave function(s) (TDSE) and the Kohn-Sham orbitals (TDDFT), such that users can build complex workflows and analyses alongside the simulations. For instance, \lstinline[basicstyle=\ttfamily\normalsize] |QMol-grid| time propagators enable arbitrary user-defined functions to be evaluated, and their result stored, while the TDSE/TDDFT propagation is performed. The package also provides built-in facilities for the calculation of common observable, including the dipole signal, energy, ionization, TDSE wave function and TDDFT Kohn-Sham orbitals. 
Aside from research purposes, the package offers a valuable resource for teaching purposes: with it, students can be introduced to a range of quantum mechanical simulation techniques (see below), using calculation examples that run on personal computers or laptops.

The \lstinline[basicstyle=\ttfamily\normalsize] |QMol-grid| package provides a suite of routines for performing quantum-mechanical simulations in atomic and molecular systems, currently implemented in one spatial dimension. 
Obviously, such lower-dimensional models cannot capture the entire manifold of processes at play in full-dimension simulations. Instead, these models play an important and complementary role in providing prototypical systems where general, non-system specific, properties can be established. A second advantage of dimensionally-reduced simulations is that they typically run at a fraction of the time of their full-dimension counterparts. This computational up-speed can then be re-invested in extended parameter scans or scouting for outcome of interest in a large parameter space. For instance, we have used this latter approach in recent analyses of ultrafast migration of charges in molecules~\cite{Mauger_2022}.
The specifics of what is included and left out in any given lower-dimension simulation is highly system/model dependent. We defer to end-users of the package to address those limitations in their specific situation.

All simulations in the \lstinline[basicstyle=\ttfamily\normalsize] |QMol-grid| package use an underlying Cartesian-grid discretization scheme, with all spatial derivatives calculated with fast-Fourier transforms. The package is written using MATLAB's object-oriented features and handle classes. Notably, the package supports:
\begin{itemize}
    \item DFT: Ground- and excited-state density-functional theory.
    \item HF: Ground- and excited-state Hartree Fock.
    \item SE: Ground- and excited-state Schrödinger equation.
    \item TDDFT: Real-time time-dependent density-functional theory.
    \item TDSE: Time-dependent Schrödinger equation.
\end{itemize} 
Ground- and excited-state calculations support both using a Cartesian grid or basis-set discretization while time-dependent simulations are currently limited to Cartesian grids.

We refer readers to the documentation for details regarding each supported computational framework. 
Briefly, within \lstinline[basicstyle=\ttfamily\normalsize] |QMol-grid|, SE provides a single-active electron model of the electronic structure of atoms and molecules. 
For multi-electron systems, HF gives the best approximation (lowest energy) of the wave function in terms of a single antisymmetrized product of one-electron wave functions (Slater determinant)~\cite{Szabo_1996}. 
Alternatively, DFT trades the multi-electron wave-function picture for the real-space electron density, whose dimension is independent of the number of active electrons. There, electron-electron interactions are captured in the (nonlinear) functional dependency of the DFT Hamiltonian on the electron density.
Specifically, \lstinline[basicstyle=\ttfamily\normalsize] |QMol-grid| uses Kohn-Sham DFT~\cite{Kohn_1965}, where the density is build from virtually-independent electrons. 
Both HF and DFT correspond to solving a nonlinear eigen-state problem, which is implemented via standard iterative techniques in the package~\cite{Johnson_1988}.

TDSE and TDDFT describe the time evolution of the system, typically either resulting from an external driving laser field or starting from a non-stationary initial state, within their respective SE and DFT framework. From its origin in ultrafast AMO science research, the \lstinline[basicstyle=\ttfamily\normalsize] |QMol-grid| package offers efficient and high-order time propagation schemes specially designed for those simulations~\cite{Mauger_2024}.
Time-dependent simulations neglect nuclear dynamics (Born-Oppenheimer approximation), with all atomic and molecular potentials fixed in space throughout the time evolution of the electrons.

%% Software description
\section{Software description}

A full description of the \lstinline[basicstyle=\ttfamily\normalsize] |QMol-grid| package, including all possible input parameters and calculation features is included in the MATLAB documentation provided with the package. After installation, the package documentation is accessible in MATLAB, in the ``Supplemental Software'' section. A copy of the documentation is also provided on the GitHub wiki.
The documentation includes a series of tutorials, starting with SE ground-state calculations, and going through TDSE, DFT, and TDDFT calculations to help new users getting familiarized with setting up calculations, input parameters, and output variables.
Throughout, the documentation also includes many script samples illustrating how one can use the various features.
Finally, the documentation discusses the required class structure for advanced users who wish to add their own functionalities to the package and inherit common interface methods to the \lstinline[basicstyle=\ttfamily\normalsize] |QMol-grid| package.

%% Software architecture
\subsection{Software architecture}

\begin{figure}[htb]
    \includegraphics[width=.8\linewidth]{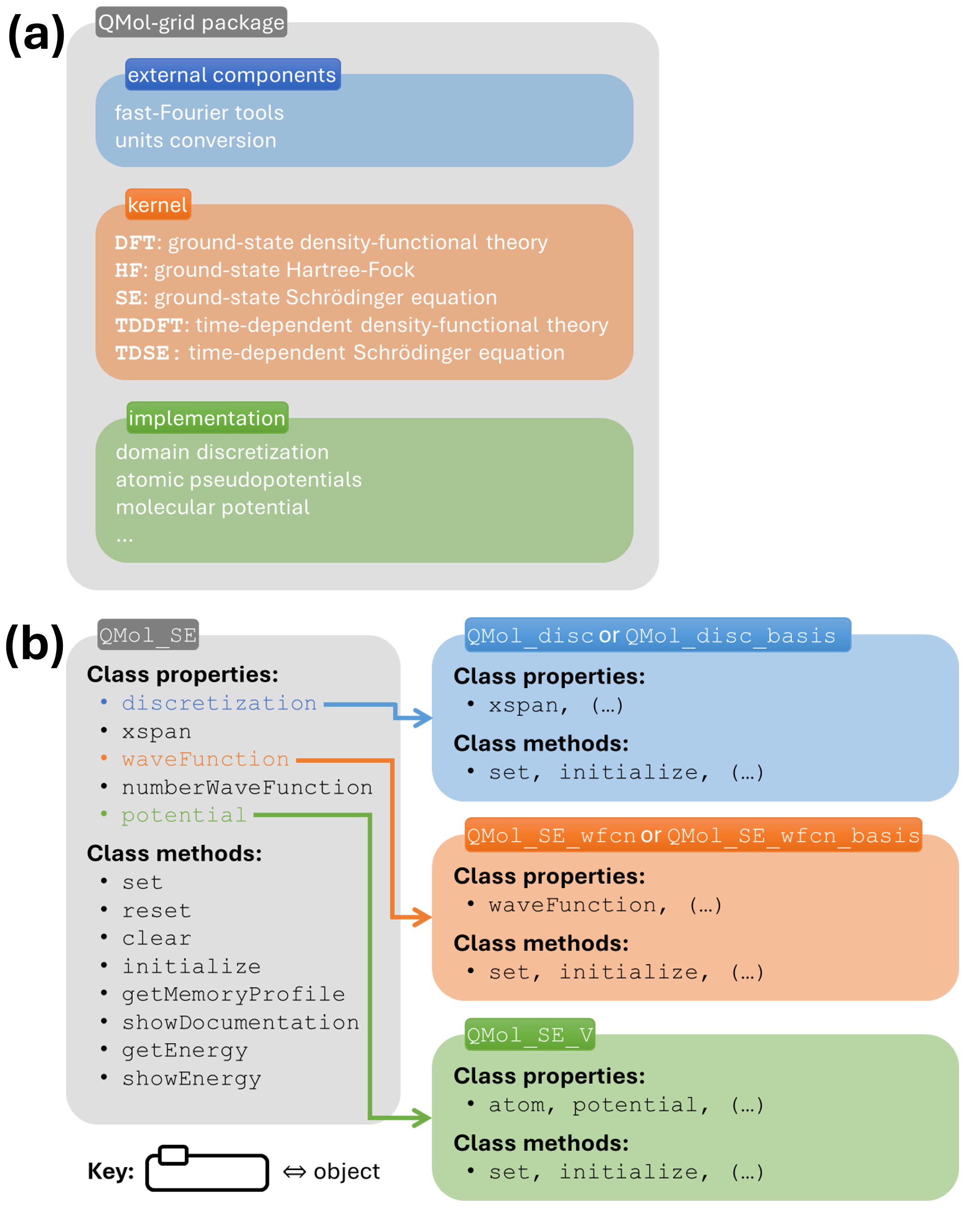}
    \centering
    \caption{
    (a) Overall architecture for the \lstinline[basicstyle=\ttfamily\normalsize] |QMol-grid| package: components are sorted in three tiers of handle classes that define the computation ecosystem.
    (b) Schematic of a Schr\"{o}dinger-equation object and its components. Each box indicates a separate class defined within the package.}
    \label{fig:architecture}
\end{figure}

The \lstinline[basicstyle=\ttfamily\normalsize] |QMol-grid| package provides an ecosystem of MATLAB handle classes. While the package is provided as a stand-alone suite, it is developed around 3 main groups sketched in figure~\ref{fig:architecture}~(a): (1) external components, (2) kernel classes that define high-level calculation methods, and (3) implementation classes that define all the lower-level functionalities.
The package is developed with the general goal of facilitating access to the wave function(s) (SE/TDSE) and the Kohn-Sham orbitals (DFT/TDDFT), which are packaged into classes for abstract manipulations of the objects in ground-state, time propagation, and common observables' calculations.

Users set up calculations by creating \lstinline[basicstyle=\ttfamily\normalsize] |QMol-grid| objects of the relevant type and specifying the desired parameters using MATLAB's common name-value pair argument structure (in arbitrary order and case insensitive). As illustrated in the examples of section~\ref{sec:examples} below, we strive to give intuitive and descriptive parameter names. The documentation provides the list, together with supported formats, of all available input parameters for each class.
Throughout the package, input parameters and output results are specified in atomic units; we provide units conversion external components to facilitate conversions to more conventional units (\textit{e.g.}, as/fs for time, W/cm$^2$ for field intensity, etc.).
Some high-level components are themselves encapsulated into classes, enabling abstract manipulations in the property objects. Figure~\ref{fig:architecture}~(b) illustrates this concept for the Schr\"{o}dinger-equation object \lstinline[basicstyle=\ttfamily\normalsize] |QMol_SE| for which class properties are a mix of variables (\lstinline[basicstyle=\ttfamily\normalsize] |xspan| and \lstinline[basicstyle=\ttfamily\normalsize] |numberWaveFunction|) and \lstinline[basicstyle=\ttfamily\normalsize] |QMol-grid| objects (\lstinline[basicstyle=\ttfamily\normalsize] |discretization|, \lstinline[basicstyle=\ttfamily\normalsize] |waveFunction|, and \lstinline[basicstyle=\ttfamily\normalsize] |potential|).
Parameters can be updated after an object has been created using the \lstinline[basicstyle=\ttfamily\normalsize] |set| method, again using name-value pair arguments.

%% Software functionalities
\subsection{Software functionalities}

Ground- and excited-state calculations in the \lstinline[basicstyle=\ttfamily\normalsize] |QMol-grid| package are performed by a direct diagonalization of the Hamiltonian operator, via MATLAB's \lstinline[basicstyle=\ttfamily\normalsize] |eigs| (grid discretization) or \lstinline[basicstyle=\ttfamily\normalsize] |eig| (basis set) functions. DFT and HF self-consistent-field iterations are performed using an Anderson's mixing scheme~\cite{Anderson_1965,Johnson_1988}.
HF is obtained by running DFT with an exact-exchange and no correlation functionals.

The time-propagators in the \lstinline[basicstyle=\ttfamily\normalsize] |QMol-grid| package are computed using symplectic-split operators~\cite{Mauger_2024} (2$^{\rm nd}$ order Strang \textit{a.k.a.} Verlet~\cite{Strang_1968}, 4$^{\rm th}$ order Forest-Ruth~\cite{Forest_1990}, and Blanes and
Moan optimized 4$^{\rm th}$ and 6${\rm th}$ order~\cite{Blanes_2002} in time, and spectral in space). They support field-free and laser-driven simulations in the dipole approximation with the following on-the-fly features, each specifying their own time sampling:
\begin{itemize}
    \item Checkpointing, with the creation of a restart MATLAB file (\lstinline[basicstyle=\ttfamily\normalsize] |.mat|) that can be used to resume a calculation that was stopped before it was finished;
    \item Calculation and storage of the dipole, dipole velocity, and dipole acceleration signals;
    \item Calculation and storage of the wave function(s)/Kohn-Sham orbitals and Hamiltonian-component energies;
    \item Storage of the wave function(s) (TDSE), and the Kohn-Sham orbitals and one-body density (TDDFT);
    \item Calculation and storage of the ionization signal, keeping track of how much electronic density is absorbed at the domain boundaries;
    \item Calculation and storage of the results of installable output functions of the wave function(s) (TDSE), and the Kohn-Sham orbitals or one-body density (TDDFT);
    \item Saving the intermediate Schrödinger- or DFT-model objects in separate MATLAB files (\lstinline[basicstyle=\ttfamily\normalsize] |.mat|).
\end{itemize}
Aside from the options that generate MATLAB files (first and last items above), the results for all the other on-the-fly calculations are collected and stored in the time propagator object itself -- see the TDDFT example in section~\ref{sec:example_2}. 
The size of the generated output strongly depends on the simulation parameters: time-dependent dipole, energy, and ionization signals are proportional to the number of saved time steps while wave functions, Kohn-Sham orbitals, and densities scale as the number of time steps multiplied by the domain grid size.
Anecdotally, in our experience dipole, energy, and ionization signals typically require a few hundred KB while saving the wave function or density easily takes a few to many MB.

Both ground/excited-state and time-propagation calculations provide run-time documentation features, providing a summary of the model and simulation configuration as well as relevant references. The run-time documentation can be toggled on (default) or off.
Profilers are also available to estimate the memory footprint and average execution time for the Hamiltonian-operator and its components.
For time-dependent simulations, the profilers provide an estimate of the size for all the on-the-fly results calculated and saved during the propagation -- see the TDDFT example in section~\ref{sec:example_2}.

The \lstinline[basicstyle=\ttfamily\normalsize] |QMol-grid| package comes with a suite of unit tests, individually checking the methods in each of the classes in the package.

%% Illustrative examples
\section{Illustrative examples} \label{sec:examples}

We illustrate how users interface with the \lstinline[basicstyle=\ttfamily\normalsize] |QMol-grid| package in two examples. 
The documentation includes a more comprehensive series of tutorials meant to get new users familiarized with how to set simulations up, interact, and recover results from calculations. Starting from ground-state SE and moving towards TDDFT, the tutorials progressively introduce (i) minimal-code examples and (ii) discussions of various input parameters and output variables available in the package.

% Schrodinger-equation ground state
\subsection{Example 1: Schr\"{o}dinger-equation ground state}

Here we illustrate how to use the \lstinline[basicstyle=\ttfamily\normalsize] |QMol-grid| package to calculate the ground-state wave function of a one-dimensional hydrogen-like atom. The Schr\"odinger-equation ground-state corresponds to the lowest-energy solution to the eigenvalue problem $\hat{\mathcal{H}}\psi(x)=E\psi(x)$, where $\hat{\mathcal{H}}$ is the Schr\"odinger-equation Hamiltonian operator, $\psi$ is the wave function, and $E$ its associated energy. In atomic units, the Hamiltonian operator is $\hat{\mathcal{H}} = -\frac{\Delta}{2} + \hat{\mathcal{V}}$.

Specifically, this example walks through defining (i) the domain and grid discretization over which the Schr\"odinger-equation and wave function are calculated, (ii) the atomic potential and (iii) the Schr\"odinger-equation model, and (iv) calculating the ground state associated with these properties.

We model the one-dimensional hydrogen model atom using a soft-Coulomb potential $V(x)=-1/\sqrt{x^2+a^2}$ with
\begin{lstlisting}[language=Octave]
H = QMol_Va_softCoulomb('softeningParameter',sqrt(2));
\end{lstlisting}
where \lstinline[basicstyle=\ttfamily\normalsize] |'softeningParameter'| specifies the value for the parameter $a$. Here we choose the softening parameter $a=\sqrt{2}$ to match H's ground state energy. By default, the atom is located at the origin $x=0$. Note that H only corresponds to the atomic model, which is shared with molecular systems and various quantum frameworks. Thus, it must be turned into a valid Schr\"odinger-equation potential, using
\begin{lstlisting}[language=Octave]
V = QMol_SE_V('atom',H);
\end{lstlisting}
Here \lstinline[basicstyle=\ttfamily\normalsize] |'atom'| indicates to the \lstinline[basicstyle=\ttfamily\normalsize] |QMol_SE_V| object that the list of atomic centers is provided next -- here a single H effective potential.

\begin{figure}[htb]
    \includegraphics[width=0.6\textwidth]{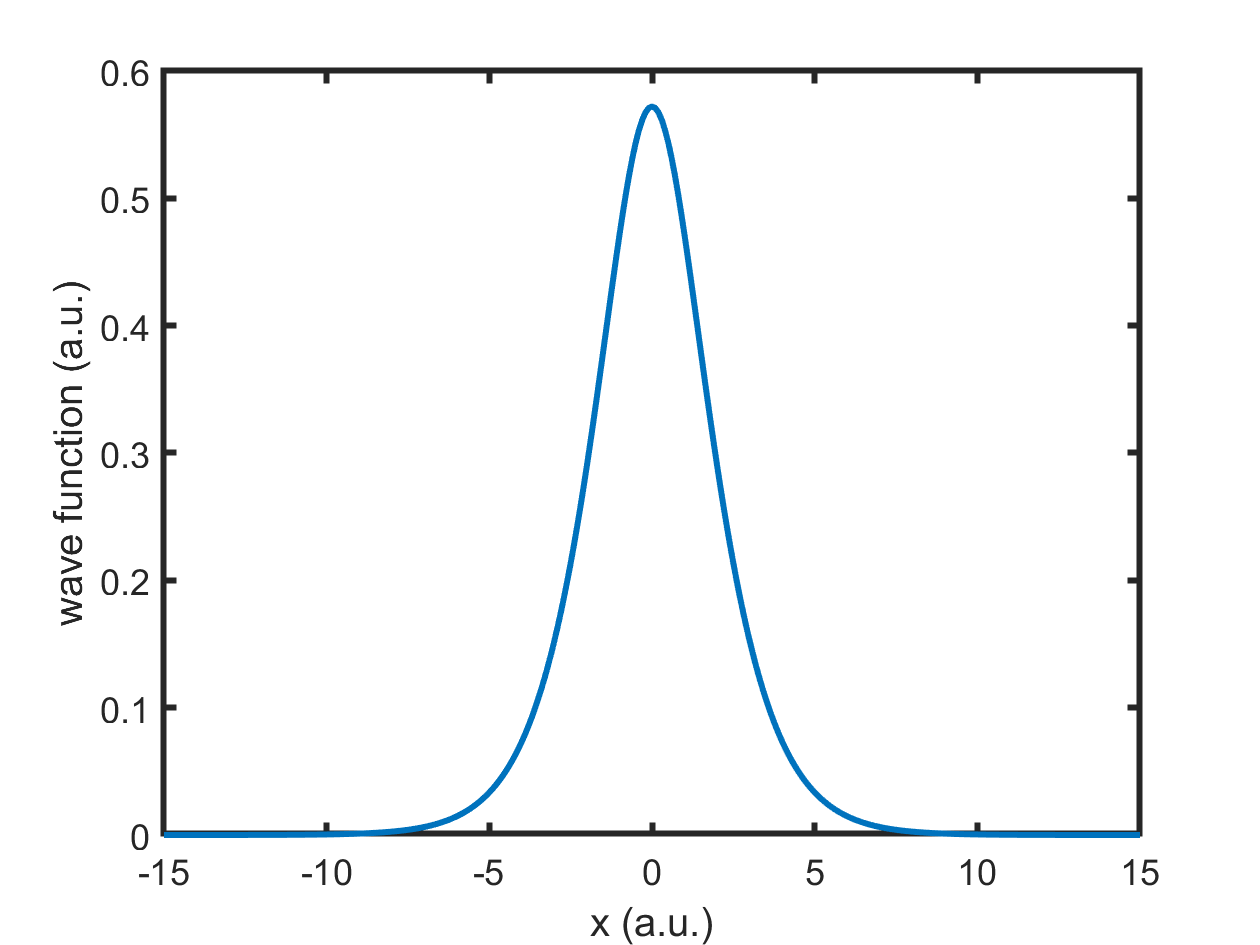}
    \centering
    \caption{Ground-state wave function $\psi(x)$ for the soft-Coulomb potential $V(x)=-1/\sqrt{x^2+2}$.}
    \label{fig:example1}
\end{figure}

The simulation domain must be a Cartesian grid -- with all increasing, equally spaced discretization points -- and should be wide enough and with small enough of a discretization step to properly capture the wave function. In our case, we select a domain ranging from -15 to 15 a.u., with a discretization steps of 0.1~a.u.
\begin{lstlisting}[language=Octave]
x = -15:.1:15;
\end{lstlisting}
We now have all the elements to define a Schr\"odinger-equation model object with the potential and domain defined above
\begin{lstlisting}[language=Octave]
SE = QMol_SE(            ...
        'xspan',     x,  ...
        'potential', V);
\end{lstlisting}
Like above, when creating the \lstinline[basicstyle=\ttfamily\normalsize] |SE| object, we recognize the definition of the discretization domain and effective potential with the keywords \lstinline[basicstyle=\ttfamily\normalsize] |'xspan'| and \lstinline[basicstyle=\ttfamily\normalsize] |'potential'|, respectively.
Next we move to calculating its associated ground-state wave function and energy using the two commands
\begin{lstlisting}[language=Octave]
GSS = QMol_SE_eigs;
GSS.computeGroundState(SE);
\end{lstlisting}
The first line creates the eigen-state solver while the second performs the actual ground-state calculation on the Schr\"odinger-equation object \lstinline[basicstyle=\ttfamily\normalsize] |SE|.
At the end of the calculation, the ground-state wave function is stored in the input \lstinline[basicstyle=\ttfamily\normalsize] |SE|, together with relevant information such as the domain discretization. For instance, solely relying on \lstinline[basicstyle=\ttfamily\normalsize] |SE|, one can plot the ground-state wave function with
\begin{lstlisting}[language=Octave]
figure
    plot(SE.xspan,SE.waveFunction.waveFunction,'-','LineWidth',2)
    set(gca,'box','on','FontSize',12,'LineWidth',2)
    xlabel('x (a.u.)')
    ylabel('wave function (a.u.)')
    xlim(SE.xspan([1 end]))
\end{lstlisting}
The output is represented in Fig.~\ref{fig:example1}. From the plot command line, we see that the domain-discretization grid may be recovered using the \lstinline[basicstyle=\ttfamily\normalsize] |xspan| property in the object \lstinline[basicstyle=\ttfamily\normalsize] |SE| (using the standard object-oriented dot notation \lstinline[basicstyle=\ttfamily\normalsize] |SE.xspan|). On the other hand, the wave function is nested inside another object, which explains the consecutive dots \lstinline[basicstyle=\ttfamily\normalsize] |SE.waveFunction.waveFunction|.
Other properties in the object \lstinline[basicstyle=\ttfamily\normalsize] |SE.waveFunction| are used by ground/excited-state and TDSE calculations; we refer to the \lstinline[basicstyle=\ttfamily\normalsize] |QMol_SE_wfcn| documentation page for further details.

% TDDFT
\subsection{Example 2: Time-dependent density-functional theory} \label{sec:example_2}

For a given set of initial Kohn-Sham orbitals, the TDDFT dynamics is described by the nonlinear system of partial differential equations, in atomic units~(a.u.)
\begin{equation}\label{eq:TDDFT}
    i\partial_t \phi_k({\bf x};t) =\hat{\mathcal{H}}_{\rm DFT}[\{\phi_k\}_k;t]({\bf x};t)\ \phi_k({\bf x};t),
\end{equation}
where $\hat{\mathcal{H}}_{\rm DFT}$ is the DFT Hamiltonian operator, which nonlinearly depends on the Kohn-Sham orbitals $\{\phi_k\}_k$.

The \lstinline[basicstyle=\ttfamily\normalsize] |QMol-grid| package relies on the canonical Hamiltonian structure of TDDFT~\cite{Mauger_2024} to integrate the dynamics of equation~(\ref{eq:TDDFT}). In this example, we illustrate how to use the \lstinline[basicstyle=\ttfamily\normalsize] |QMol-grid| package to integrate the TDDFT dynamics of an open-shell one-dimensional molecular ion model with 3 atomic centers and 5 active electrons.

\textit{Initial condition:} 
In the \lstinline[basicstyle=\ttfamily\normalsize] |QMol-grid| package, TDDFT simulations are decoupled from setting up the initial condition, which must be done independently.
Similar to example 1, we build the molecular model out of 3 one-dimensional atomic models, each contributing 2 electrons to the molecule, using soft-Coulomb potentials.
For our example, we start by calculating the neutral-molecule ground state:
\begin{lstlisting}[language=Octave]
% Molecular model
V_1     =   QMol_Va_softCoulomb( ...
                'atom','X_1','charge',2,'position',-3);
V_2     =   QMol_Va_softCoulomb( ...
                'atom','X_2','charge',2,'position', 0);
V_3     =   QMol_Va_softCoulomb( ...
                'atom','X_3','charge',2,'position', 3);

% DFT model
Vext    =   QMol_DFT_Vext('atom',{V_1,V_2,V_3});
Vh      =   QMol_DFT_Vh_conv;
Vxc     =  {QMol_DFT_Vx_LDA_soft,QMol_DFT_Vc_LDA_soft};

DFT     =   QMol_DFT_spinPol(                                  ...
                'xspan',                       -50:.1:50,      ...
                'occupation',               {[1 1 1],[1 1 1]}, ...
                'externalPotential',            Vext,          ...
                'HartreePotential',             Vh,            ...
                'exchangeCorrelationPotential', Vxc,           ...
                'selfInteractionCorrection',    'ADSIC'        );

% DFT ground state
SCF     =   QMol_DFT_SCF_Anderson;
SCF.solveSCF(DFT);
\end{lstlisting}
The ``\lstinline[basicstyle=\ttfamily\normalsize] |% Molecular model|'' block defines the atomic effective potential, specifying the name, bare charge, and location of each atomic center, respectively.
The ``\lstinline[basicstyle=\ttfamily\normalsize] |% DFT model|'' block first defines the molecular potential \lstinline[basicstyle=\ttfamily\normalsize] |Vext|, followed by the DFT functionals \lstinline[basicstyle=\ttfamily\normalsize] |Vh| and \lstinline[basicstyle=\ttfamily\normalsize] |Vxc| to be used in the (TD)DFT calculations -- see the documentation's ground-state DFT tutorial for further details regarding the model parameters. 
The final block ``\lstinline[basicstyle=\ttfamily\normalsize] |% DFT ground state|'' first creates the eigen-state DFT solver, here an Anderson mixing scheme~\cite{Johnson_1988}, and performs the ground-state self-consistent field (SCF) calculation.

Next, we manually induce an excitation in the molecular cation by successively (i) replacing one of the Kohn-sham orbitals by a superposition of molecular-orbital states (excitation part) and (ii) removing an electron, going from 3 to 2, from the down-spin Kohn-Sham orbitals (ionization part).
\begin{lstlisting}[language=Octave]
% Induce excitation
DFT.orbital.set('orbitalDown', [DFT.KSO.KSOdw(:,1)  ...
   (DFT.KSO.KSOdw(:,2)+DFT.KSO.KSOdw(:,3))/sqrt(2)]);

% Induce ionization
DFT.set('occupation',{[1 1 1],[1 1]});
\end{lstlisting}
We now have a non-stationary set of Kohn-Sham orbitals, leading to field-free dynamics under equation~(\ref{eq:TDDFT}).

\textit{TDDFT simulation:}
With the DFT molecular model and the initial condition in hand, we now move to integrating the subsequent field-free TDDFT dynamics. For this, we select a fourth-order Forest-Ruth symplectic split-operator scheme~\cite{Forest_1990,Mauger_2024}.
Note that, here the field-free TDDFT dynamics does not lead to any ionization and therefore no boundary conditions need be specified at the edges of the domain. For field-driven simulations, absorbing boundary conditions can be specified to avoid spurious boundary effects.
\begin{lstlisting}[language=Octave]
TDDFT   =   QMol_TDDFT_SSO_4FR(                     ...
                'time',                 0:10:100,   ...
                'timeStep',             2e-2,       ...
                'saveDensity',          true,       ...
                'saveDensityTime',      1);
\end{lstlisting}
In our example, the TDDFT object is created with:
\begin{itemize}
    \item The first pair of arguments specifies that the integration should start at time t=0 and end at t=100 a.u. The step of 10 a.u., is unrelated to the propagation time step and instead specifies the time intervals to use in the progress display.
    \item The second pair of arguments specifies the (fixed) time step for the propagation.
    \item The third pair of arguments indicates that the one-body density should be saved periodically, with the period specified by the fourth pair of arguments, \textit{i.e.}, every 1~a.u. in our case.
\end{itemize}
Then, we launch the TDDFT integration with
\begin{lstlisting}[language=Octave]
TDDFT.propagate(DFT);
\end{lstlisting}
At the end of the simulation, the DFT object has been updated to contain the Kohn-Sham orbitals at $t=100$~a.u. The time-dependent one-body density is stored in the TDDFT object itself.

\begin{figure}[htb]
    \includegraphics[width=0.6\textwidth]{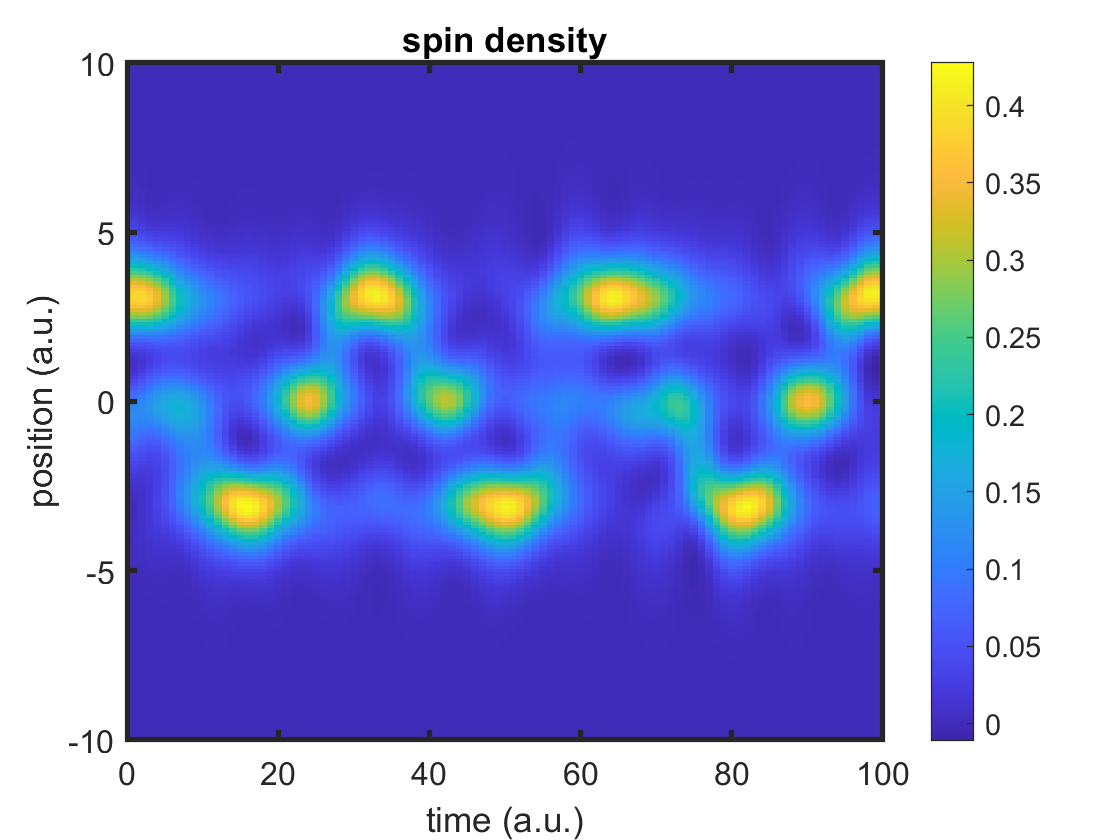}
    \centering
    \caption{Evolution of the spin density, defined as the difference between the up- and down-spin one-body densities, along the molecular model we consider for our TDDFT-simulation example.}
    \label{fig:example2}
\end{figure}

\textit{Plotting the result:}
Next we recover calculated observables out of the TDDFT object. Each set of observable is stored in a separate structure property in the TDDFT object, which contains (i) the exact time vector at which the quantity has been saved and (ii) the observable itself. In our case, the structure of interest is \lstinline[basicstyle=\ttfamily\normalsize] |TDDFT.outDensity| with the up- and down-spin densities respectively stored in the fields \lstinline[basicstyle=\ttfamily\normalsize] |totalUp| and \lstinline[basicstyle=\ttfamily\normalsize] |totalDown|. The densities are matrices with columns corresponding to the successive saved times. To plot the spin density, defined as the difference between the up- and down-spin one-body densities, we use
\begin{lstlisting}[language=Octave]
figure
    imagesc(TDDFT.outDensity.time,DFT.xspan, ...
        TDDFT.outDensity.totalUp-TDDFT.outDensity.totalDown)
    set(gca,'box','on','FontSize',12,'LineWidth',2,'YDir','normal')
    xlim(TDDFT.outDensity.time([1 end]))
    ylim([-10 10])
    xlabel('time (a.u.)')
    ylabel('position (a.u.)')
    title('spin density')
    colorbar vert
\end{lstlisting}
with the result shown in Fig.~\ref{fig:example2}.

\textit{Profiling (estimating the memory footprint):}
Before running the TDDFT calculation, users have the possibility to check how much memory the simulation requires to run and store the requested one-body densities. Using the same calculation workflow as above, right after creating the \lstinline[basicstyle=\ttfamily\normalsize] |TDDFT| object, the memory footprint is obtained with
\begin{lstlisting}[language=Octave]
TDDFT.initialize(DFT);
QMol_DFT_profiler(TDDFT,'memory');
\end{lstlisting}
In our case, the estimated total TDDFT-object size is 1.8~MB with 1.5~MB for the saved electron density. 
Saving the \lstinline[basicstyle=\ttfamily\normalsize] |TDDFT| and \lstinline[basicstyle=\ttfamily\normalsize] |DFT| object in a MATLAB file at the end of the propagation produces a 1.6~MB \lstinline[basicstyle=\ttfamily\normalsize] |.mat| file. We mostly attribute the slight difference with the profiler estimate to run-time memory overhead associated with internal variables that are not stored in the saved objects.

%% Impact
\section{Impact}

The \lstinline[basicstyle=\ttfamily\normalsize] |QMol-grid| package offers a versatile suite of quantum simulation techniques for reduced-dimension atomic and molecular models. Its native MATLAB structure facilitates on-the-fly calculations and analyses in time-dependent simulations as well as post-processing, which all can be done using high-level functionalities of MATLAB. Simulation data are organized within handle classes with common interface methods to simplify end-user interaction with the various components of the package.
\lstinline[basicstyle=\ttfamily\normalsize] |QMol-grid| comes with a full documentation, including many script samples that illustrate how one can use the various features. It also includes a series of tutorials to guide new users with setting up calculations, input parameters, and output variables.

In our groups, we used an early development version of the \lstinline[basicstyle=\ttfamily\normalsize] |QMol-grid| package in~\cite{Mauger_2022} for nonlinear analysis of ultrafast migration of electronic charges in molecules. Notably, the efficacy of simulations allowed us to perform thousands of TDDFT simulations and with it get a detailed picture of the migration-dynamics phase space, something that is essentially unfeasible in full-dimension quantum packages.
More recently, we used \lstinline[basicstyle=\ttfamily\normalsize] |QMol-grid| to validate symplectic split-operator propagation schemes for TDDFT~\cite{Mauger_2024}. The symplectic propagators (4$^{\rm th}$ order Forest-Ruth~\cite{Forest_1990}, and Blanes and
Moan optimized 4$^{\rm th}$ and 6${\rm th}$ order~\cite{Blanes_2002} schemes) are now integrated and available in the package -- see example~2 of section~\ref{sec:example_2}.
We continue to use \lstinline[basicstyle=\ttfamily\normalsize] |QMol-grid| in various on-going projects in our groups.
Outside of a research environment, the package could be used for teaching: thanks to the modest computational requirements, students could run illustrative examples of quantum mechanics or (TD)DFT on personal computers or laptops.

%% Conclusions
\section{Conclusions}

The \lstinline[basicstyle=\ttfamily\normalsize] |QMol-grid| package provides a versatile suite of quantum-mechanical methods at the Schr\"{o}dinger, Hartree-Fock, and density-functional theory levels of theory for ground- and excited-state calculations, as well as TDSE and TDDFT propagators. Time-propagation schemes provide streamlined access to the wave function(s) (TDSE) and the Kohn-Sham orbitals (TDDFT).
The wave functions and Kohn-Sham orbitals are packaged into classes that enable abstract manipulations in the objects, \textit{e.g.}, for ground-state, time propagation, and common observables' calculations.
The object-oriented structure provides a uniform user interface, where input parameters are specified as pairs of parameter-name/parameter-value (in arbitrary order and case insensitive). 
Output results are stored in the objects and can be recovered using standard object-oriented dot notation -- see the tutorials for examples.

%% CREDITS
\section*{CRediT author statement}

\textbf{F. Mauger:} Conceptualization, Software, Validation, Documentation, Writing - Original Draft, Funding acquisition.
\textbf{C. Chandre:} Documentation, Writing - Original Draft.

%% Acknowledgements
\section*{Acknowledgements}

FM thanks M.B.~Gaarde, K.~Lopata, and K.J.~Schafer for enlightening discussions, suggestions, and support throughout the development of the package.
The original development of the \lstinline[basicstyle=\ttfamily\normalsize] |QMol-grid| package, and its (TD)DFT features, was supported by the U.S. Department of Energy, Office of Science, Office of Basic Energy Sciences, under Award No.~DE-SC0012462. 
Addition of the (TD)SE features was supported by the National Science Foundation under Grant No.~PHY-2207656

%%%%%%%%%%%%%%%%%%%%%%%%%%%%%%%%%%%%%%%%%%%%%%%%%%%%%%%%%%%%%
%%%%%%%%%%%%%%%%%%%%%%%%%%%%%%%%%%%%%%%%%%%%%%%%%%%%%%%%%%%%%
%%                       BIBLIOGRAPHY                      %%
%%%%%%%%%%%%%%%%%%%%%%%%%%%%%%%%%%%%%%%%%%%%%%%%%%%%%%%%%%%%%
%%%%%%%%%%%%%%%%%%%%%%%%%%%%%%%%%%%%%%%%%%%%%%%%%%%%%%%%%%%%%
%\bibliographystyle{unsrt}
%\bibliography{bibliography} 

%\textit{If the software repository you used supplied a DOI or another
%Persistent IDentifier (PID), please add a reference for your software
%here. For more guidance on software citation, please see our guide for
%authors or \href{https://f1000research.com/articles/9-1257/v2}{this
%  article on the essentials of software citation by FORCE 11}, of
%which Elsevier is a member.}

\end{document}